\documentclass[twocolumn,pra,superscriptaddress,showpacs,aps]{revtex4-1}
\usepackage{color}
\usepackage{amsmath}
\usepackage{amssymb}
\usepackage{txfonts}
\usepackage{graphicx}
\usepackage[unicode]{hyperref}
\hypersetup{pdftitle={EIT cavity},
 pdfauthor={S-W. Su, WenTe Liao},
 pdfsubject={EIT cavity},
  colorlinks=true,
    linkcolor=blue,
    citecolor=blue,
    filecolor=black,
    urlcolor=blue}



\begin{document}

\title{All-optical control of superradiance and matter waves using a dispersive cavity}

\author{Shih-Wei Su}

\affiliation{Department of Physics and Graduate Institute of Photonics, National
Changhua University of Education, Changhua 50058 Taiwan}

\author{Zhen-Kai Lu}

\affiliation{Max Planck Institute for Quantum Optics, D-85748 Garching, Germany}

\author{Nina Rohringer}

\affiliation{Max Planck Institute for the Structure and Dynamics of Matter, 22761
Hamburg, Germany}

\affiliation{Max Planck Institute for the Physics of Complex Systems, 01187 Dresden,
Germany}

\affiliation{Center for Free-Electron Laser Science, 22761 Hamburg, Germany}

\author{Shih-Chuan Gou}

\affiliation{Department of Physics and Graduate Institute of Photonics, National
Changhua University of Education, Changhua 50058 Taiwan}

\author{Wen-Te Liao}

\email{wenteliao@cc.ncu.edu.tw}


\affiliation{Max Planck Institute for the Structure and Dynamics of Matter, 22761
Hamburg, Germany}

\affiliation{Max Planck Institute for the Physics of Complex Systems, 01187 Dresden,
Germany}

\affiliation{Center for Free-Electron Laser Science, 22761 Hamburg, Germany}

\affiliation{Department of Physics, National Central University, 32001 Taoyuan
City, Taiwan}

\begin{abstract}
Cavity quantum electrodynamics (CQED) \cite{haroche1997b}
plays an elegant role of studying strong coupling between light and
matter. However, a non-mechanical, direct and dynamical control of
the used mirrors is still unavailable. Here we theoretically investigate
a novel type of dynamically controllable cavity composed of two atomic
mirrors. Based on the electromagnetically induced transparency (EIT)
\cite{kocharovskaya1986,boller1991}, the
reflectance of atomic mirror is highly controllable through its dispersive
properties by varying the intensity of applied coupling fields or
the optical depth of atomic media. To demonstrate the uniqueness of
the present cavity, we further show the possibility of manipulating
vacuum-induced diffraction of a binary Bose-Einstein condensate (BEC)
\cite{Ketterle2002,brennecke2007} when loading it into a dispersive
cavity and experiencing superradiant scatterings \cite{ketterle1999}.
Our results may provide a novel all-optical element for atom optics \cite{meystre2001} and shine new light on controlling
light-matter interaction. 
\end{abstract}

\date{Version of \today. }

\maketitle

For decades, CQED has served as an elegant model for demonstrating atom-light interaction and fundamental principles of quantum mechanics \cite{haroche1997b}. In conjunction with the state-of-art experiments using ultracold atoms, CQED has been able to facilitate the realization of quantum metrology \cite{Ian2010}, quantum teleportation \cite{Duan2001} and quantum memory storage \cite{Kimble2008}, etc., experiments paving the way for the fulfillment of quantum technology.  By its very nature, cavity is an indispensable ingredient to CQED, which reflects photons and has them interact with atoms in a confined space. However, conventional CQED systems consisting of reflecting mirrors limiting themselves from being networked at large scale. Here, we propose a novel cavity system where mechanical mirrors are replaced by atomic ensembles whose dispersion can be controlled by, e.g., EIT, and so we term it as dispersive cavity which offers unique controllability.
Moreover, the separation between the atomic mirrors and their size can be much smaller than that of mechanical ones. Therefore, the cavity cooperativity is enhanced due to smaller cavity volume and it becomes more feasible to reach strong coupling regime and  scalable for more demanding systems. 
Both advantages render fabricating a more functional matter-wave interferometry or atomic circuit possible.
The dispersive cavity can also be integrated  on atom chips \cite{schumm2005} or used to study the cavity-mediated spin-orbit coupling in the ultracold atoms~\cite{Dong2014}, bringing both the controllability and the tiny size into full play.


\begin{figure*}[tbh]
\includegraphics[width=0.6\paperwidth]{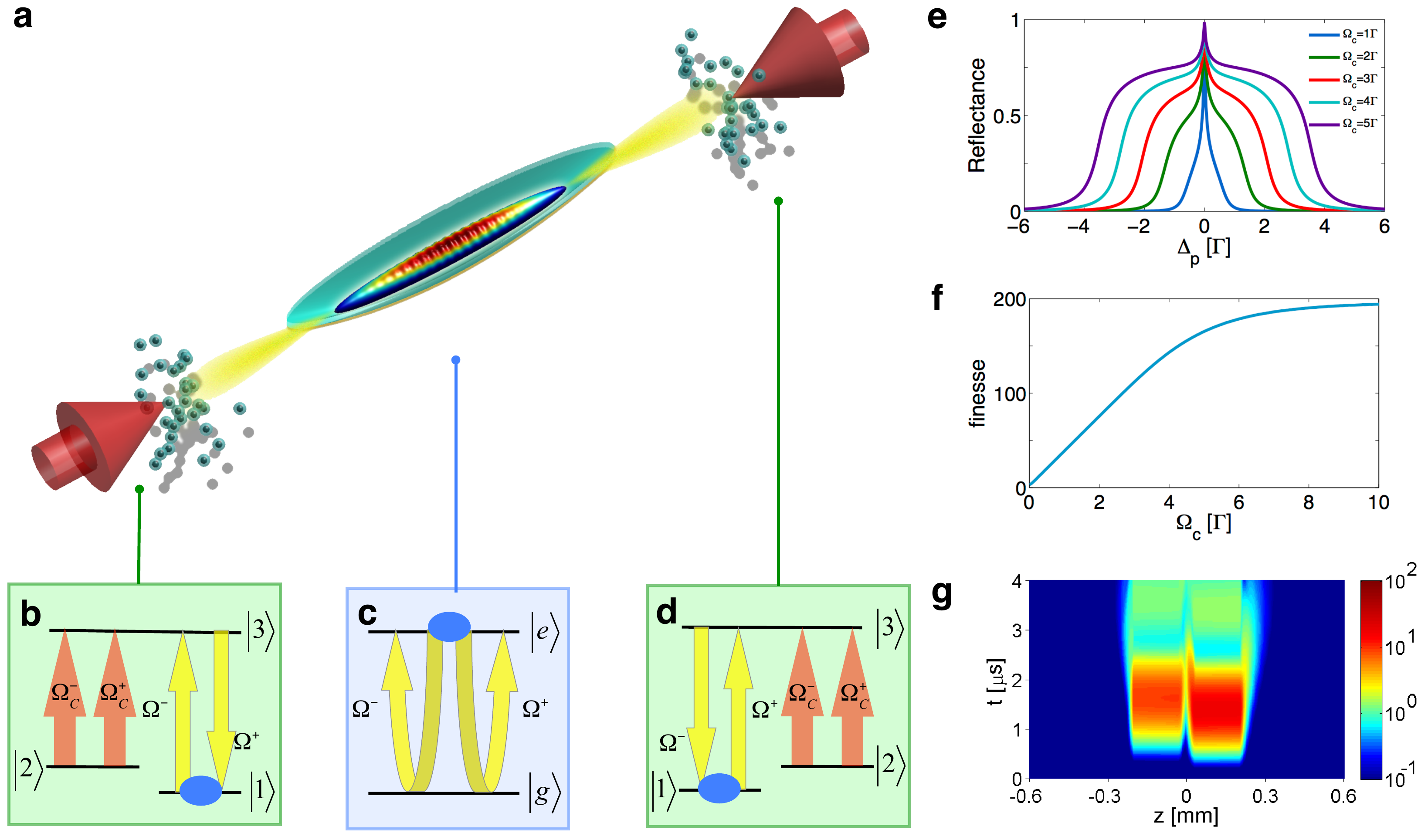}
\caption{\textbf{Schematic plot of a dispersive cavity.} \textbf{a}, a binary BEC sandwiched by two three-level EIT media, where the former is initially in the state $\left|e\right\rangle $ and the latters are 
in state $\left|1\right\rangle$. The red
arrows represent two counter-propagating coupling fields driving the transition $\left|2\right\rangle \rightarrow\left|3\right\rangle $
as shown in \textbf{b} and \textbf{d}. 
The quantum fluctuations of binary BEC spontaneously generate
the forward and backward SR fields $\Omega^{\pm}$ (yellow) as shown in \textbf{c}, which connect the  EIT  transition from $\left|1\right\rangle \rightarrow\left|3\right\rangle $.  
The composition of three-level atoms and coupling fields acts as a dispersive cavity,  which reflects resonant yellow photons. 
The controllable reflectance of a dispersive mirror and the enhancement of cavity finesse are shown in \textbf{e} and \textbf{f}, respectively. \textbf{g}, the coherently trapped superradiant field $[\vert\Omega^{+}(t,z)\vert^2+\vert\Omega^{-}(t,z)\vert^2]/\Gamma^2$ is shown where EIT mirrors are located in the regions $|z|>0.2$ mm. }
\label{system} 
\end{figure*}
\begin{figure*}[tbh]
\includegraphics[width=0.6\paperwidth]{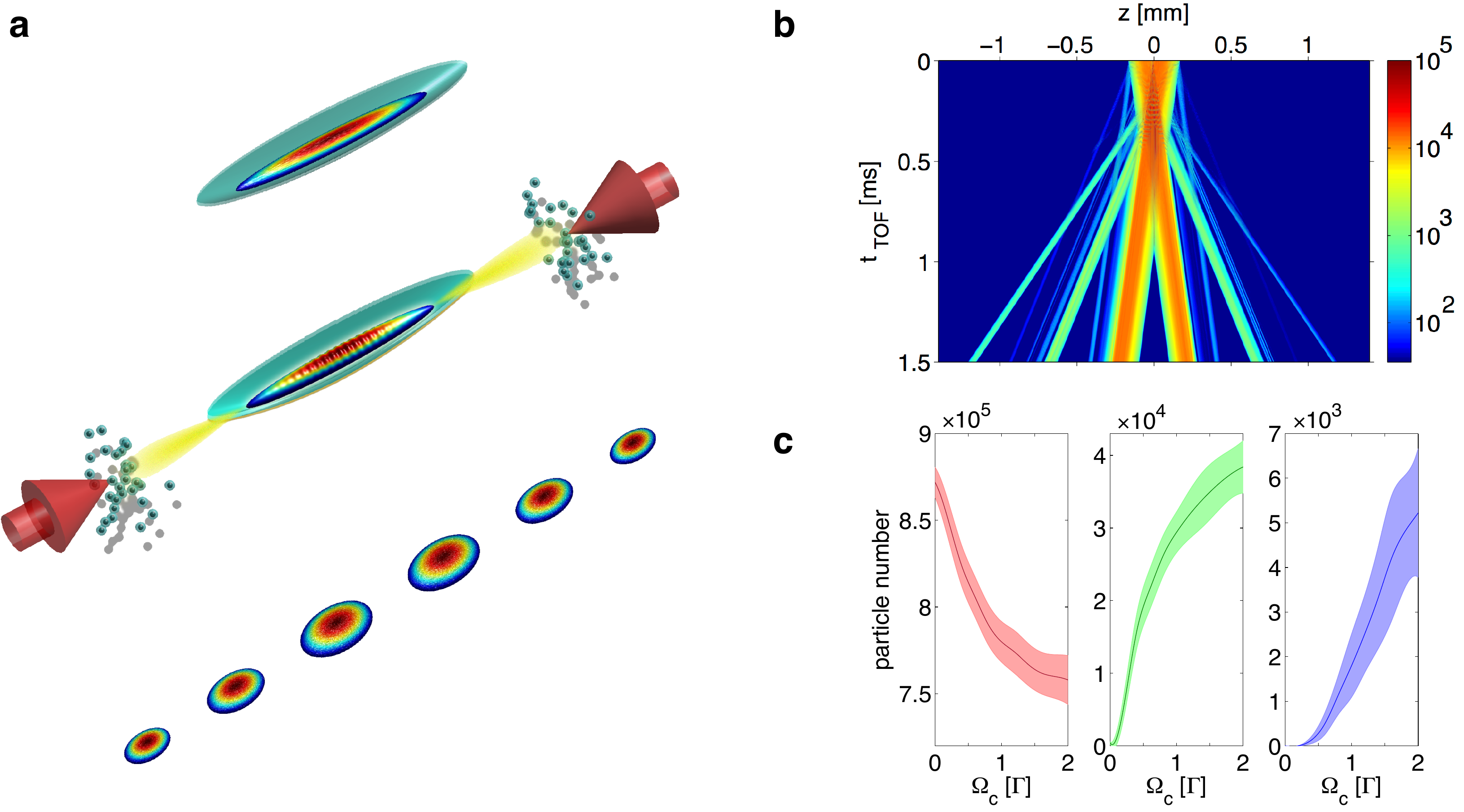}
\caption{\textbf{Controllable vacuum induced diffraction of matter waves using a dispersive cavity.} 
\textbf{a}, coherent matter waves initially  prepared in its internal excited state are diffracted by three steps from top to bottom. The condensate is created in a harmonic trap (green surface) and flies through a dispersive cavity where odd multiple $k_p$ modes of matter waves emerge. When switching off the trap, matter waves of different modes split due to the momentum kick from SR photons, namely, atomic diffraction. 
\textbf{b}, the BEC's time of flight dynamics. Six density bumps carrying momenta of $\pm k_p$, $\pm 3k_p$, and $\pm 5k_p$ can be clearly observed at the end of time.
\textbf{c}, the coupling-field-strength-dependent mean particle number of different modes of diffracted BEC carrying $\pm k_{p}$ (red), $\pm 3k_{p}$ (green) and $\pm 5k_{p}$ (blue).
All the data points are averaged over $1000$ realizations, and the color filled regions illustrate the corresponding error bars.
}
\label{single-trajectory-r0-3um-OD-500-Om-3} 
\end{figure*}

\begin{figure*}[tbh]
\includegraphics[width=0.6\paperwidth]{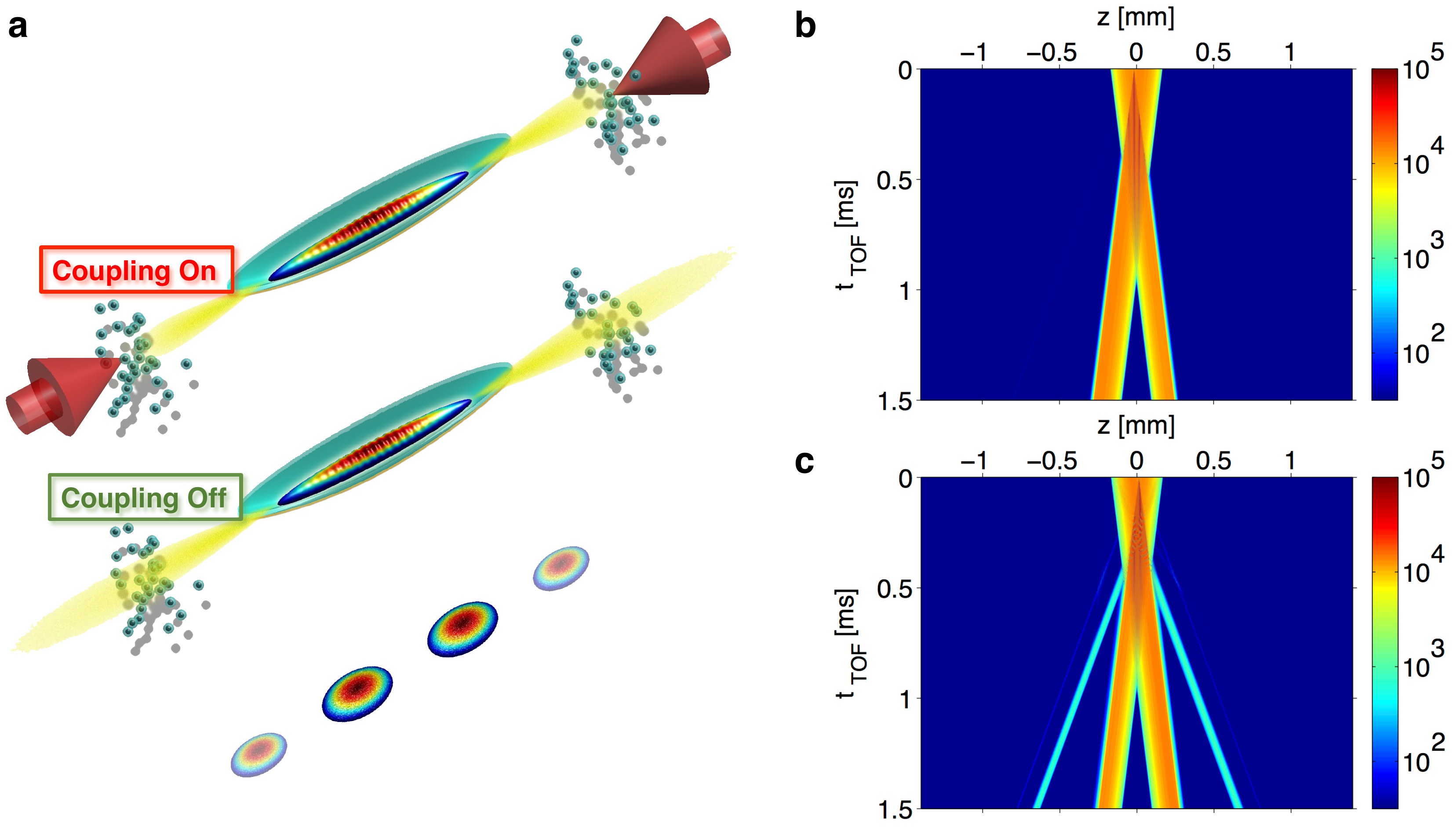}
\caption{\textbf{Dynamical control of vacuum induced diffraction of matter waves using a dispersive cavity.}
\textbf{a}, when loading BEC into a dispersive cavity, atomic diffraction can be dynamically controlled by turning off the coupling field at different instance. 
The time of flight for that coupling field is off at \textbf{b} 0.02 $\mu$s and \textbf{c} 0.07 $\mu$s after loading BEC. The particle numbers on each mode of \textbf{b} and \textbf{c}  are $(N_{\pm k_{p}},N_{\pm 3k_{p}},N_{\pm 5k_{p}})=$$(8.6\times 10^{5},300,0)$ and $(8.2\times 10^5, 1.6\times 10^4,200)$, respectively.  
 }
\label{Fig_3} 
\end{figure*}

As depicted in Fig.~\ref{system} {\bf a}, our dispersive cavity is composed
of two remote collections of three-level-$\Lambda$ type atoms. 
Based on EIT \cite{bajcsy2003,lin2009},
as illustrated in Fig.~\ref{system} {\bf b} and {\bf d},
each atomic medium constitutes a mirror when interacting with
two counter-propagating coupling fields with Rabi frequencies $\Omega_{c}^{\pm}$
driving the $\left|2\right\rangle \leftrightarrow \left|3\right\rangle $ transition. Under the action of four-wave mixing \cite{lin2009},
an incident probe field with Rabi frequency $\Omega^{+}$ driving
the transition $\left|3\right\rangle \leftrightarrow\left|1\right\rangle $
will be reflected and generate
a backward probe field with Rabi frequency $\Omega^{-}$ \cite{bajcsy2003}, and vise versa.
For showing the controllability and novelty of the dispersive cavity, we
consider an excited quasi one-dimensional binary
BEC 
which transversely flies through
it and emits photons \cite{brennecke2007}. As illustrated in Fig.~\ref{system} \textbf{c}, the binary BEC consists of atoms condensed
in the ground state $\left|g\right\rangle $ and excited state $\left|e\right\rangle $.
The photonic polarization and frequency of transition $\left|g\right\rangle \rightarrow\left|e\right\rangle $
match that of $\left|1\right\rangle \rightarrow\left|3\right\rangle $
of two atomic mirrors such that emitted photons with wavelength $\lambda$ from BEC can be reflected by two side EIT media. 
The vacuum fluctuations \cite{glauber1979}
and the geometric shape of the BEC will speed up the decay and result in the emission of
superradiance (SR) along BEC's long axis \cite{glauber1979,haroche1982,ketterle1999}.
Therefore when the long axis of the loaded BEC 
is parallel to the cavity axis, the SR will circulate within a dispersive
cavity and become subsequently releasable by switching off coupling fields. 
The coexistence of BEC and superradiant photons within a cavity enables the multiple scattering between them. This renders the generation of matter waves with high order harmonic of photon wave number possible, namely, the diffraction of matter waves \cite{meystre2001}. Because the effect is triggered by vacuum fluctuations, we term it as vacuum induced diffraction of matter waves.

We first investigate
the controllability of the novel cavity and calculate the reflectance
of an EIT mirror by numerically solving
the optical-Bloch
equations \cite{lin2009,Thorsten2012} 
\begin{eqnarray}
 &  & \partial_{t}\hat{\rho}=\frac{1}{i\hbar}\left[\hat{H},\hat{\rho}\right]+\hat{\rho}_{dec},\label{eq1}\\
 &  & \left(\frac{1}{c}\partial_{t}\pm\partial_{z}\right)\Omega^{\pm}=i\eta_{EIT}\rho_{31}^{\pm},\label{eq2}
\end{eqnarray}
where $\hat{\rho}$ is the density matrix for the state vector
$\sum_{i=1}^{3}B_{i}\vert i\rangle$ of the three-level atom in the Fig.~\ref{system}~\textbf{a, b} and \textbf{d}. Also, $\hat{\rho}_{dec}$
describes the spontaneous decay of the excited state $\left|3\right\rangle $
characterized by rate $\Gamma$. $\hat{H}$ is the Hamiltonian
for the interaction between atomic mirrors and counter-propagating
fields (see Methods). Moreover, $\eta_{EIT}=\Gamma d^{opt}/2L$
with $d^{opt}$ and $L$ being the optical depth and the length of
the medium, respectively. 
The resultant dispersive cavity has only one reflecting band whose
width is associated to the EIT transparency window $\Delta\omega_{EIT}=\Omega_{c}^{2}/\left(\Gamma\sqrt{d^{opt}}\right)$
which is highly tunable by either varying the optical depth or the coupling fields.
As shown
in Fig.~\ref{system} \textbf{e}, at various detuning and a fixed optical depth $d^{opt}=500$, the reflectance approaches unity when the
fields $\Omega^{\pm}$ are resonant and the width can be controlled
by varying the intensity of the coupling fields. 
To gain more insight
into the  dispersive cavity, we invoke
the effective cavity finesse $f(\Delta_p)=\frac{\pi}{1-R(\Delta_p)}$
where $R(\Delta_p)$ is the reflectance as a function of the detuning
$\Delta_p$ of fields $\Omega^{\pm}$ \cite{chang2012}. 
The cavity finesse is plotted in Fig.~\ref{system} \textbf{f}, which demonstrates
that the cavity finesse can be controlled by tuning the EIT transparency window \cite{kocharovskaya1986,boller1991}.

We are ready to show the cavity
effect on the BEC dynamics by numerically integrating equations~(\ref{eq1}) and (\ref{eq2})
together with
the coupled Gross-Pitaevskii equation which reads \cite{ginsberg2007,Zhang2009} (see methods)
\begin{eqnarray}
i\hbar\partial_{t}\left(\begin{array}{c}
\psi_{g}\\
\psi_{e}
\end{array}\right) & = & \left(\begin{array}{cc}
H_{sp}+H_{int}^{g} & H_{atom-light}\\
H_{atom-light}^{*} & H_{sp}+H_{int}^{e}-\frac{i\Gamma}{2}
\end{array}\right)\left(\begin{array}{c}
\psi_{g}\\
\psi_{e}
\end{array}\right),\label{eq:GPE}
\end{eqnarray}
where $\psi_{g,e}$ are the wave functions of the binary BEC,
$H_{sp}$ the single-particle Hamiltonian, $H_{int}^{g,e}$ the atom-atom
interaction, $H_{atom-light}$ the atom-light interaction (see Methods),
and $\Gamma$ is the spontaneous decay rate of the excited level $\left|e\right\rangle $
introduced phenomenologically. 
To start up the superradiant
process of the binary BEC in a cavity, the vacuum fluctuation is introduced
by truncated Wigner method \cite{Blakie2008} (see
methods). The dynamics of the generated SR
fields is described by
\begin{align}
\left(\frac{1}{c}\partial_{t}\pm\partial_{z}\right)\Omega^{\pm} & =i\eta_{BEC}\sigma_{eg}^{(\pm1)},\label{eq:Maxwell-Eq-BEC}
\end{align}
where $\sigma_{eg}^{\left(\pm1\right)}$ is the coherence
of the BEC associated with the plane-wave factor $e^{\pm ik_{p}z}$ (see Methods for details)
and $\eta_{BEC}=3\Gamma N_{BEC}\lambda^{2}/4\pi A$ with $A$ being
the transverse cross section of the condensate. In the following, we consider the EIT mirror of optical depth $d^{opt}=500$ and coupling field of $\Omega_c=2\Gamma$.
A quasi one-dimentional  BEC of $^7$Li at  2S$_{1/2}$ ($\left|g\right\rangle $) state can be  prepared in a single beam dipole trap. An additional  TEM$_{01}$-type laser mode beam is applied to cross the atomic ensemble \cite{esslinger2012}. The spatial intensity distribution of the cross beam allows to trap low density  thermal clouds aside as the EIT mirrors while leaving the central BEC undisturbed.  The $^7$Li BEC can be coherently transferred to the excited 3P$_{3/2}$ ($\left|e\right\rangle $)  state by a $\pi$ pulse. The narrow linewidth of 3P$_{3/2}$ state, $\Gamma\sim 760$ kHz \cite{hulet2011} offers a reasonable time scale for detection.   Moreover, $d^{opt}=1000$ has already been experimentally achieved in cold atom systems \cite{hsiao2014,blatt2014}.
As an example, with $N_{BEC}=3\times10^{5}$ and $A=(\pi\times3^{2})$ $\mu$m$^2$, Fig.~\ref{system} \textbf{g} depicts the intensity of the confined  superradiance  $\vert\Omega^{+}(t,z)\vert^2+\vert\Omega^{-}(t,z)\vert^2$ in a dispersive cavity.

The three steps of vacuum induced diffraction of matter waves is shown in Fig.~\ref{single-trajectory-r0-3um-OD-500-Om-3} \textbf{a}, namely,  from top to bottom, creating BEC in state $\left|e\right\rangle $, loading BEC into a dispersive cavity and releasing BEC.
The multiple-scattering between circulating SR fields
and the BEC generates higher order harmonics of the matter wave. 
Theses  high $k_{p}$ modes 
can be clearly observed by time-of-flight (TOF) simulation with
$N_{BEC}=10^{6}$
 as shown in Fig.~\ref{single-trajectory-r0-3um-OD-500-Om-3} \textbf{b}. Especially at $\mathrm{t_{TOF}}=1.5$ ms there are six notable peaks corresponding to $-5 k_{p}$, $-3 k_{p}$, $-k_{p}$,  $k_{p}$, $3 k_{p}$ and $5 k_{p}$ from left to right. 
To study the controllability of generating high-$k_{p}$ modes, we tune the cavity finesse by varying
the strength of coupling fields and perform $1000$ independent realizations for statistics. As shown
in Fig.~\ref{single-trajectory-r0-3um-OD-500-Om-3} \textbf{c}, the population of $\pm k_{p}$ modes is gradually and coherently transferred to that of higher order harmonics while increasing
$\Omega_{c}$.
The particle number of $\pm3k_{p}$
and $\pm5k_{p}$ modes are significantly generated when $\Omega_{c}\gtrsim 0.1\Gamma$
and $1\Gamma$, respectively.  
This is in contrast to the situation
without the dispersive cavity where only $\pm k_{p}$ modes and very small fraction of $\pm 3 k_{p}$ modes are generated, namely, about a factor
of 1000 enhancement by the cavity for $\pm3k_{p}$ modes. Remarkably, $\pm5k_{p}$ modes only emerge under the action of a dispersive cavity.

Figure~\ref{Fig_3} illustrates the possibility of dynamically manipulating atomic diffraction. In Fig.~\ref{Fig_3} {\bf a} we demonstrate a protocol to suppress the generation of particular harmonic of $k_{p}$ matter wave by switching off two coupling fields at some instance, which dynamically disables the dispersive cavity. We respectively turn off the coupling fields at 0.02 $\mu$s and 0.07 $\mu$s in Fig.~\ref{Fig_3} {\bf b} and Fig.~\ref{Fig_3} {\bf c} after BEC is loaded. Our TOF simulations show that $\pm3k_{p}$
and $\pm5k_{p}$ modes can be successfully suppressed by release SR photons at proper instances before corresponding multiple scattering happens.

In conclusion, the novel flexibility of a dispersive cavity may give the possibility of implementing
more delicate manipulations of matter wave
and  becomes all-optical elements for atom optics. 
Additionally, an all-optical  Q-switching superradiant source may result from combining our system with triggered superradiance \cite{keitel1992}.
Moreover, the recent advance on x-ray quantum optics \cite{rohringer2012,cavaletto2014} also shows the potential of achieving atomic mirrors in x-ray domain \cite{wang2015}. Given the high, non-mechanical
and dynamical controllability of the dispersive cavity , the present system provides a unique and novel stage of manipulating light-mater interaction.

\section*{Methods}
In the presence of the counter-propagating coupling and SR fields,
the Hamiltonian of the EIT medium is $\hat{H}=\hat{H_{0}}+\hat{H_{I}}$ where 
$\hat{H_{0}}=\sum_{j}\hbar\omega_{j}\left|j\right\rangle\left\langle j\right|$ and $\hat{H_{I}}$ is
the atom-light interaction Hamiltonian of the EIT medium given by
form as 
\begin{eqnarray*}
\hat{H_{I}} & = &\frac{\hbar}{2} \left[\left(\Omega^{+}e^{i(k_{p}z-\omega_{p}t)}+\Omega^{-}e^{i(-k_{p}z-\omega_{p}t)}\right)\left|3\right\rangle \left\langle 1\right|\right.\\
 &  & \left.+\left(\Omega_{c}^{+}e^{i(k_{c}z-\omega_{c}t)}+\Omega_{c}^{-}e^{i(-k_{c}z-\omega_{c}t)}\right)\left|3\right\rangle \left\langle 2\right|+H.c.\right].
\end{eqnarray*}

In equation~(\ref{eq:GPE}), the operators accounting for the quasi 1D condensate dynamics take the form as
$H_{sp}=-\frac{\hbar^{2}\partial_{z}^{2}}{2m}+\frac{1}{2}m\omega_{z}^{2}z^{2}$, $H_{atom-light}=\frac{\hbar}{2}\left(\Omega^{+*}e^{-ik_{p}z}+\Omega^{-*}e^{ik_{p}z}\right)$
and $H_{int}^{j}=g_{jj}\left|\psi_{l}\right|^{2}+g_{jl}\left|\psi_{l}\right|^{2}$ which are the single-particle, atom-light interaction, and nonlinear interaction Hamiltonians, respectively. The quasi 1D nonlnear interaction constant takes the form as $g_{jl}=\frac{2\pi a_{jl}\omega_{r}N_{bec}}{\omega_{z}a_{osc}}$ where $a_{osc}=\sqrt{\frac{\hbar}{m\omega_{z}}}$ is oscillator length. The $\omega_{r}$ and $\omega_{z}$ are the transverse and longitudinal trapping frequencies, respectively, where we use  $(\omega_{r},\omega_{z})=2\pi\times(100,2.1\times10^{3})$~Hz in the simulation. The optical coherence of the condensate is given by $\sigma_{eg}=\psi_{e}\psi_{g}^{*}$. In the presence of circulating SR fields, the macroscopic wave functions and coherence would consist of high $\pm nk_p$ modes, i.e. $\psi_g=\sum_{n=-\infty}^{\infty}\psi_{g}^{(n)}e^{i(2n+1)k_{p}z}$, $\psi_e=\sum_{n=-\infty}^{\infty}\psi_{e}^{(n)}e^{i2nk_{p}z}$, and $\sigma_{eg}=\sum_{n=-\infty}^{\infty}\sigma_{eg}^{(n)}e^{ink_{p}z}$ with $\psi_{e,g}^{(n)}$ and $\sigma_{eg}^{(n)}$ being the spatially slowly-varying envelope. The coherence terms, $\sigma_{eg}^{(\pm 1)}$ carrying $e^{\pm ik_{p}z}$ account for the generation of SR fields.

The dynamics of the condensate flying through the cavity is simulated by the initial state where
all the atoms are condensed in the excited state forming a Thomas-Fermi density profile while the ground state consists of quantum fluctuations sampled according
to the truncated Wigner approximation of $\psi_{g}=\sqrt{1/N_{BEC}}\sum_{j=1}^{M}\alpha_{j}\xi_{j}\left(z\right)$,
 $<\alpha_{j}^{*}\alpha_{l}>=\delta_{jl}/2$, and $\xi_{j}\left(z\right)$ being
an orthonormal basis. 

Equations~(\ref{eq1})-(\ref{eq:Maxwell-Eq-BEC}) are numerically solved by the method of lines where the Bloch and GP equations are propagated by the Fourier pseudospectral method
and the adaptive Runge-Kutta method of orders 4 and 5 (RK45) for space and time integration, respectively, while the SR fields are integrated by the semi-Euler method. In the simulation, the total number of Fourier modes is $5\times2^{12}$ and the number of mode for sampling initial fluctuations is $M=3000$.


\begin{thebibliography}{10}%
\makeatletter
\providecommand \@ifxundefined [1]{%
 \@ifx{#1\undefined}
}%
\providecommand \@ifnum [1]{%
 \ifnum #1\expandafter \@firstoftwo
 \else \expandafter \@secondoftwo
 \fi
}%
\providecommand \@ifx [1]{%
 \ifx #1\expandafter \@firstoftwo
 \else \expandafter \@secondoftwo
 \fi
}%
\providecommand \natexlab [1]{#1}%
\providecommand \enquote  [1]{``#1''}%
\providecommand \bibnamefont  [1]{#1}%
\providecommand \bibfnamefont [1]{#1}%
\providecommand \citenamefont [1]{#1}%
\providecommand \href@noop [0]{\@secondoftwo}%
\providecommand \href [0]{\begingroup \@sanitize@url \@href}%
\providecommand \@href[1]{\@@startlink{#1}\@@href}%
\providecommand \@@href[1]{\endgroup#1\@@endlink}%
\providecommand \@sanitize@url [0]{\catcode `\\12\catcode `\$12\catcode
  `\&12\catcode `\#12\catcode `\^12\catcode `\_12\catcode `\%12\relax}%
\providecommand \@@startlink[1]{}%
\providecommand \@@endlink[0]{}%
\providecommand \url  [0]{\begingroup\@sanitize@url \@url }%
\providecommand \@url [1]{\endgroup\@href {#1}{\urlprefix }}%
\providecommand \urlprefix  [0]{URL }%
\providecommand \Eprint [0]{\href }%
\providecommand \doibase [0]{http://dx.doi.org/}%
\providecommand \selectlanguage [0]{\@gobble}%
\providecommand \bibinfo  [0]{\@secondoftwo}%
\providecommand \bibfield  [0]{\@secondoftwo}%
\providecommand \translation [1]{[#1]}%
\providecommand \BibitemOpen [0]{}%
\providecommand \bibitemStop [0]{}%
\providecommand \bibitemNoStop [0]{.\EOS\space}%
\providecommand \EOS [0]{\spacefactor3000\relax}%
\providecommand \BibitemShut  [1]{\csname bibitem#1\endcsname}%
\let\auto@bib@innerbib\@empty
\bibitem [{\citenamefont {Ma\^{i}tre}\ \emph {et~al.}(1997)\citenamefont
  {Ma\^{i}tre}, \citenamefont {Hagley}, \citenamefont {Nogues}, \citenamefont
  {Wunderlich}, \citenamefont {Goy}, \citenamefont {Brune}, \citenamefont
  {Raimond},\ and\ \citenamefont {Haroche}}]{haroche1997b}%
  \BibitemOpen
  \bibfield  {author} {\bibinfo {author} {\bibfnamefont {X.}~\bibnamefont
  {Ma\^{i}tre}}, \bibinfo {author} {\bibfnamefont {E.}~\bibnamefont {Hagley}},
  \bibinfo {author} {\bibfnamefont {G.}~\bibnamefont {Nogues}}, \bibinfo
  {author} {\bibfnamefont {C.}~\bibnamefont {Wunderlich}}, \bibinfo {author}
  {\bibfnamefont {P.}~\bibnamefont {Goy}}, \bibinfo {author} {\bibfnamefont
  {M.}~\bibnamefont {Brune}}, \bibinfo {author} {\bibfnamefont {J.~M.}\
  \bibnamefont {Raimond}}, \ and\ \bibinfo {author} {\bibfnamefont
  {S.}~\bibnamefont {Haroche}},\ }\href@noop {} {\bibfield  {journal} {\bibinfo
   {journal} {Phys. Rev. Lett.}\ }\textbf {\bibinfo {volume} {79}},\ \bibinfo
  {pages} {769} (\bibinfo {year} {1997})}\BibitemShut {NoStop}%
\bibitem [{\citenamefont {Kocharovskaya}\ and\ \citenamefont
  {Khanin}(1986)}]{kocharovskaya1986}%
  \BibitemOpen
  \bibfield  {author} {\bibinfo {author} {\bibfnamefont {O.}~\bibnamefont
  {Kocharovskaya}}\ and\ \bibinfo {author} {\bibfnamefont {Y.~I.}\ \bibnamefont
  {Khanin}},\ }\href@noop {} {\bibfield  {journal} {\bibinfo  {journal} {Sov.
  Phys. JETP}\ }\textbf {\bibinfo {volume} {63}},\ \bibinfo {pages} {945}
  (\bibinfo {year} {1986})}\BibitemShut {NoStop}%
\bibitem [{\citenamefont {Boller}\ \emph {et~al.}(1991)\citenamefont {Boller},
  \citenamefont {Imamo\ifmmode~\breve{g}\else \u{g}\fi{}lu},\ and\
  \citenamefont {Harris}}]{boller1991}%
  \BibitemOpen
  \bibfield  {author} {\bibinfo {author} {\bibfnamefont {K.-J.}\ \bibnamefont
  {Boller}}, \bibinfo {author} {\bibfnamefont {A.}~\bibnamefont
  {Imamo\ifmmode~\breve{g}\else \u{g}\fi{}lu}}, \ and\ \bibinfo {author}
  {\bibfnamefont {S.~E.}\ \bibnamefont {Harris}},\ }\href@noop {} {\bibfield
  {journal} {\bibinfo  {journal} {Phys. Rev. Lett.}\ }\textbf {\bibinfo
  {volume} {66}},\ \bibinfo {pages} {2593} (\bibinfo {year}
  {1991})}\BibitemShut {NoStop}%
\bibitem [{\citenamefont {Ketterle}(2002)}]{Ketterle2002}%
  \BibitemOpen
  \bibfield  {author} {\bibinfo {author} {\bibfnamefont {W.}~\bibnamefont
  {Ketterle}},\ }\href@noop {} {\bibfield  {journal} {\bibinfo  {journal} {Rev.
  Mod. Phys.}\ }\textbf {\bibinfo {volume} {74}},\ \bibinfo {pages} {1131}
  (\bibinfo {year} {2002})}\BibitemShut {NoStop}%
\bibitem [{\citenamefont {Brennecke}\ \emph {et~al.}(2007)\citenamefont
  {Brennecke}, \citenamefont {Donner}, \citenamefont {Ritter}, \citenamefont
  {Bourdel}, \citenamefont {K{\"o}hl},\ and\ \citenamefont
  {Esslinger}}]{brennecke2007}%
  \BibitemOpen
  \bibfield  {author} {\bibinfo {author} {\bibfnamefont {F.}~\bibnamefont
  {Brennecke}}, \bibinfo {author} {\bibfnamefont {T.}~\bibnamefont {Donner}},
  \bibinfo {author} {\bibfnamefont {S.}~\bibnamefont {Ritter}}, \bibinfo
  {author} {\bibfnamefont {T.}~\bibnamefont {Bourdel}}, \bibinfo {author}
  {\bibfnamefont {M.}~\bibnamefont {K{\"o}hl}}, \ and\ \bibinfo {author}
  {\bibfnamefont {T.}~\bibnamefont {Esslinger}},\ }\href@noop {} {\bibfield
  {journal} {\bibinfo  {journal} {Nature}\ }\textbf {\bibinfo {volume} {450}},\
  \bibinfo {pages} {268} (\bibinfo {year} {2007})}\BibitemShut {NoStop}%
\bibitem [{\citenamefont {Inouye}\ \emph {et~al.}(1999)\citenamefont {Inouye},
  \citenamefont {Chikkatur}, \citenamefont {Stamper-Kurn}, \citenamefont
  {Stenger}, \citenamefont {Pritchard},\ and\ \citenamefont
  {Ketterle}}]{ketterle1999}%
  \BibitemOpen
  \bibfield  {author} {\bibinfo {author} {\bibfnamefont {S.}~\bibnamefont
  {Inouye}}, \bibinfo {author} {\bibfnamefont {A.}~\bibnamefont {Chikkatur}},
  \bibinfo {author} {\bibfnamefont {D.}~\bibnamefont {Stamper-Kurn}}, \bibinfo
  {author} {\bibfnamefont {J.}~\bibnamefont {Stenger}}, \bibinfo {author}
  {\bibfnamefont {D.}~\bibnamefont {Pritchard}}, \ and\ \bibinfo {author}
  {\bibfnamefont {W.}~\bibnamefont {Ketterle}},\ }\href@noop {} {\bibfield
  {journal} {\bibinfo  {journal} {Science}\ }\textbf {\bibinfo {volume}
  {285}},\ \bibinfo {pages} {571} (\bibinfo {year} {1999})}\BibitemShut
  {NoStop}%
\bibitem [{\citenamefont {Meystre}()}]{meystre2001}%
  \BibitemOpen
  \bibfield  {author} {\bibinfo {author} {\bibfnamefont {P.}~\bibnamefont
  {Meystre}},\ }\href@noop {} {\emph {\bibinfo {title} {Atom optics}}}\
  (\bibinfo  {publisher} {Springer Science \& Business Media})\BibitemShut
  {NoStop}%
\bibitem [{\citenamefont {Leroux}\ \emph {et~al.}(2010)\citenamefont {Leroux},
  \citenamefont {Schleier-Smith},\ and\ \citenamefont {Vuletic}}]{Ian2010}%
  \BibitemOpen
  \bibfield  {author} {\bibinfo {author} {\bibfnamefont {I.~D.}\ \bibnamefont
  {Leroux}}, \bibinfo {author} {\bibfnamefont {M.~H.}\ \bibnamefont
  {Schleier-Smith}}, \ and\ \bibinfo {author} {\bibfnamefont {V.}~\bibnamefont
  {Vuletic}},\ }\href@noop {} {\bibfield  {journal} {\bibinfo  {journal} {Phys.
  Rev. Lett.}\ }\textbf {\bibinfo {volume} {104}} (\bibinfo {year}
  {2010})}\BibitemShut {NoStop}%
\bibitem [{\citenamefont {Duan}\ \emph {et~al.}(2001)\citenamefont {Duan},
  \citenamefont {Lukin}, \citenamefont {Cirac},\ and\ \citenamefont
  {Zoller}}]{Duan2001}%
  \BibitemOpen
  \bibfield  {author} {\bibinfo {author} {\bibfnamefont {L.-M.}\ \bibnamefont
  {Duan}}, \bibinfo {author} {\bibfnamefont {M.~D.}\ \bibnamefont {Lukin}},
  \bibinfo {author} {\bibfnamefont {J.~I.}\ \bibnamefont {Cirac}}, \ and\
  \bibinfo {author} {\bibfnamefont {P.}~\bibnamefont {Zoller}},\ }\href@noop {}
  {\bibfield  {journal} {\bibinfo  {journal} {Nature}\ }\textbf {\bibinfo
  {volume} {414}} (\bibinfo {year} {2001})}\BibitemShut {NoStop}%
\bibitem [{\citenamefont {Kimble}(2008)}]{Kimble2008}%
  \BibitemOpen
  \bibfield  {author} {\bibinfo {author} {\bibfnamefont {H.~J.}\ \bibnamefont
  {Kimble}},\ }\href@noop {} {\bibfield  {journal} {\bibinfo  {journal}
  {Nature}\ }\textbf {\bibinfo {volume} {453}} (\bibinfo {year}
  {2008})}\BibitemShut {NoStop}%
\bibitem [{\citenamefont {Schumm}\ \emph {et~al.}(2005)\citenamefont {Schumm},
  \citenamefont {Hofferberth}, \citenamefont {Andersson}, \citenamefont
  {Wildermuth}, \citenamefont {Groth}, \citenamefont {Bar-Joseph},
  \citenamefont {Schmiedmayer},\ and\ \citenamefont {Kr{\"u}ger}}]{schumm2005}%
  \BibitemOpen
  \bibfield  {author} {\bibinfo {author} {\bibfnamefont {T.}~\bibnamefont
  {Schumm}}, \bibinfo {author} {\bibfnamefont {S.}~\bibnamefont {Hofferberth}},
  \bibinfo {author} {\bibfnamefont {L.~M.}\ \bibnamefont {Andersson}}, \bibinfo
  {author} {\bibfnamefont {S.}~\bibnamefont {Wildermuth}}, \bibinfo {author}
  {\bibfnamefont {S.}~\bibnamefont {Groth}}, \bibinfo {author} {\bibfnamefont
  {I.}~\bibnamefont {Bar-Joseph}}, \bibinfo {author} {\bibfnamefont
  {J.}~\bibnamefont {Schmiedmayer}}, \ and\ \bibinfo {author} {\bibfnamefont
  {P.}~\bibnamefont {Kr{\"u}ger}},\ }\href@noop {} {\bibfield  {journal}
  {\bibinfo  {journal} {Nature Physics}\ }\textbf {\bibinfo {volume} {1}},\
  \bibinfo {pages} {57} (\bibinfo {year} {2005})}\BibitemShut {NoStop}%
\bibitem [{\citenamefont {Dong}\ \emph {et~al.}(2014)\citenamefont {Dong},
  \citenamefont {Zhou}, \citenamefont {Wu}, \citenamefont {Ramachandhran},\
  and\ \citenamefont {Pu}}]{Dong2014}%
  \BibitemOpen
  \bibfield  {author} {\bibinfo {author} {\bibfnamefont {L.}~\bibnamefont
  {Dong}}, \bibinfo {author} {\bibfnamefont {L.}~\bibnamefont {Zhou}}, \bibinfo
  {author} {\bibfnamefont {B.}~\bibnamefont {Wu}}, \bibinfo {author}
  {\bibfnamefont {B.}~\bibnamefont {Ramachandhran}}, \ and\ \bibinfo {author}
  {\bibfnamefont {H.}~\bibnamefont {Pu}},\ }\href@noop {} {\bibfield  {journal}
  {\bibinfo  {journal} {Phys. Rev. A}\ }\textbf {\bibinfo {volume} {89}},\
  \bibinfo {pages} {011602(R)} (\bibinfo {year} {2014})}\BibitemShut {NoStop}%
\bibitem [{\citenamefont {Bajcsy}\ \emph {et~al.}(2003)\citenamefont {Bajcsy},
  \citenamefont {Zibrov},\ and\ \citenamefont {Lukin}}]{bajcsy2003}%
  \BibitemOpen
  \bibfield  {author} {\bibinfo {author} {\bibfnamefont {M.}~\bibnamefont
  {Bajcsy}}, \bibinfo {author} {\bibfnamefont {A.~S.}\ \bibnamefont {Zibrov}},
  \ and\ \bibinfo {author} {\bibfnamefont {M.~D.}\ \bibnamefont {Lukin}},\
  }\href@noop {} {\bibfield  {journal} {\bibinfo  {journal} {Nature}\ }\textbf
  {\bibinfo {volume} {426}},\ \bibinfo {pages} {638} (\bibinfo {year}
  {2003})}\BibitemShut {NoStop}%
\bibitem [{\citenamefont {Lin}\ \emph {et~al.}(2009)\citenamefont {Lin},
  \citenamefont {Liao}, \citenamefont {Peters}, \citenamefont {Chou},
  \citenamefont {Wang}, \citenamefont {Cho}, \citenamefont {Kuan},\ and\
  \citenamefont {Yu}}]{lin2009}%
  \BibitemOpen
  \bibfield  {author} {\bibinfo {author} {\bibfnamefont {Y.-W.}\ \bibnamefont
  {Lin}}, \bibinfo {author} {\bibfnamefont {W.-T.}\ \bibnamefont {Liao}},
  \bibinfo {author} {\bibfnamefont {T.}~\bibnamefont {Peters}}, \bibinfo
  {author} {\bibfnamefont {H.-C.}\ \bibnamefont {Chou}}, \bibinfo {author}
  {\bibfnamefont {J.-S.}\ \bibnamefont {Wang}}, \bibinfo {author}
  {\bibfnamefont {H.-W.}\ \bibnamefont {Cho}}, \bibinfo {author} {\bibfnamefont
  {P.-C.}\ \bibnamefont {Kuan}}, \ and\ \bibinfo {author} {\bibfnamefont
  {I.~A.}\ \bibnamefont {Yu}},\ }\href@noop {} {\bibfield  {journal} {\bibinfo
  {journal} {Phys. Rev. Lett.}\ }\textbf {\bibinfo {volume} {102}},\ \bibinfo
  {pages} {213601} (\bibinfo {year} {2009})}\BibitemShut {NoStop}%
\bibitem [{\citenamefont {Haake}\ \emph {et~al.}(1979)\citenamefont {Haake},
  \citenamefont {King}, \citenamefont {Schr\"oder}, \citenamefont {Haus},\ and\
  \citenamefont {Glauber}}]{glauber1979}%
  \BibitemOpen
  \bibfield  {author} {\bibinfo {author} {\bibfnamefont {F.}~\bibnamefont
  {Haake}}, \bibinfo {author} {\bibfnamefont {H.}~\bibnamefont {King}},
  \bibinfo {author} {\bibfnamefont {G.}~\bibnamefont {Schr\"oder}}, \bibinfo
  {author} {\bibfnamefont {J.}~\bibnamefont {Haus}}, \ and\ \bibinfo {author}
  {\bibfnamefont {R.}~\bibnamefont {Glauber}},\ }\href@noop {} {\bibfield
  {journal} {\bibinfo  {journal} {Phys. Rev. A}\ }\textbf {\bibinfo {volume}
  {20}},\ \bibinfo {pages} {2047} (\bibinfo {year} {1979})}\BibitemShut
  {NoStop}%
\bibitem [{\citenamefont {Gross}\ and\ \citenamefont
  {Haroche}(1982)}]{haroche1982}%
  \BibitemOpen
  \bibfield  {author} {\bibinfo {author} {\bibfnamefont {M.}~\bibnamefont
  {Gross}}\ and\ \bibinfo {author} {\bibfnamefont {S.}~\bibnamefont
  {Haroche}},\ }\href@noop {} {\bibfield  {journal} {\bibinfo  {journal}
  {Physics Reports}\ }\textbf {\bibinfo {volume} {93}},\ \bibinfo {pages} {301}
  (\bibinfo {year} {1982})}\BibitemShut {NoStop}%
\bibitem [{\citenamefont {Peters}\ \emph {et~al.}(2012)\citenamefont {Peters},
  \citenamefont {Su}, \citenamefont {Chen}, \citenamefont {Wang}, \citenamefont
  {Gou},\ and\ \citenamefont {Yu}}]{Thorsten2012}%
  \BibitemOpen
  \bibfield  {author} {\bibinfo {author} {\bibfnamefont {T.}~\bibnamefont
  {Peters}}, \bibinfo {author} {\bibfnamefont {S.-W.}\ \bibnamefont {Su}},
  \bibinfo {author} {\bibfnamefont {Y.-H.}\ \bibnamefont {Chen}}, \bibinfo
  {author} {\bibfnamefont {J.-S.}\ \bibnamefont {Wang}}, \bibinfo {author}
  {\bibfnamefont {S.-C.}\ \bibnamefont {Gou}}, \ and\ \bibinfo {author}
  {\bibfnamefont {I.~A.}\ \bibnamefont {Yu}},\ }\href@noop {} {\bibfield
  {journal} {\bibinfo  {journal} {Phys. Rev. A}\ }\textbf {\bibinfo {volume}
  {85}},\ \bibinfo {pages} {023838} (\bibinfo {year} {2012})}\BibitemShut
  {NoStop}%
\bibitem [{\citenamefont {Chang}\ \emph {et~al.}(2012)\citenamefont {Chang},
  \citenamefont {Jiang}, \citenamefont {Gorshkov},\ and\ \citenamefont
  {Kimble}}]{chang2012}%
  \BibitemOpen
  \bibfield  {author} {\bibinfo {author} {\bibfnamefont {D.~E.}\ \bibnamefont
  {Chang}}, \bibinfo {author} {\bibfnamefont {L.}~\bibnamefont {Jiang}},
  \bibinfo {author} {\bibfnamefont {A.}~\bibnamefont {Gorshkov}}, \ and\
  \bibinfo {author} {\bibfnamefont {H.}~\bibnamefont {Kimble}},\ }\href@noop {}
  {\bibfield  {journal} {\bibinfo  {journal} {New Journal of Physics}\ }\textbf
  {\bibinfo {volume} {14}},\ \bibinfo {pages} {063003} (\bibinfo {year}
  {2012})}\BibitemShut {NoStop}%
\bibitem [{\citenamefont {Ginsberg}\ \emph {et~al.}(2007)\citenamefont
  {Ginsberg}, \citenamefont {Garner},\ and\ \citenamefont
  {Hau}}]{ginsberg2007}%
  \BibitemOpen
  \bibfield  {author} {\bibinfo {author} {\bibfnamefont {N.~S.}\ \bibnamefont
  {Ginsberg}}, \bibinfo {author} {\bibfnamefont {S.~R.}\ \bibnamefont
  {Garner}}, \ and\ \bibinfo {author} {\bibfnamefont {L.~V.}\ \bibnamefont
  {Hau}},\ }\href@noop {} {\bibfield  {journal} {\bibinfo  {journal} {Nature}\
  }\textbf {\bibinfo {volume} {445}},\ \bibinfo {pages} {623} (\bibinfo {year}
  {2007})}\BibitemShut {NoStop}%
\bibitem [{\citenamefont {Zhang}\ \emph {et~al.}(2009)\citenamefont {Zhang},
  \citenamefont {Cui}, \citenamefont {Zhou},\ and\ \citenamefont
  {Liu}}]{Zhang2009}%
  \BibitemOpen
  \bibfield  {author} {\bibinfo {author} {\bibfnamefont {J.~M.}\ \bibnamefont
  {Zhang}}, \bibinfo {author} {\bibfnamefont {F.~C.}\ \bibnamefont {Cui}},
  \bibinfo {author} {\bibfnamefont {D.~L.}\ \bibnamefont {Zhou}}, \ and\
  \bibinfo {author} {\bibfnamefont {W.~M.}\ \bibnamefont {Liu}},\ }\href@noop
  {} {\bibfield  {journal} {\bibinfo  {journal} {Phys. Rev. A}\ }\textbf
  {\bibinfo {volume} {79}},\ \bibinfo {pages} {033401} (\bibinfo {year}
  {2009})}\BibitemShut {NoStop}%
\bibitem [{\citenamefont {Blakie}\ \emph {et~al.}(2008)\citenamefont {Blakie},
  \citenamefont {Bradley}, \citenamefont {Davis}, \citenamefont {Ballagh},\
  and\ \citenamefont {Gardiner}}]{Blakie2008}%
  \BibitemOpen
  \bibfield  {author} {\bibinfo {author} {\bibfnamefont {P.~B.}\ \bibnamefont
  {Blakie}}, \bibinfo {author} {\bibfnamefont {A.~S.}\ \bibnamefont {Bradley}},
  \bibinfo {author} {\bibfnamefont {M.~J.}\ \bibnamefont {Davis}}, \bibinfo
  {author} {\bibfnamefont {R.~J.}\ \bibnamefont {Ballagh}}, \ and\ \bibinfo
  {author} {\bibfnamefont {C.~W.}\ \bibnamefont {Gardiner}},\ }\href@noop {}
  {\bibfield  {journal} {\bibinfo  {journal} {Adv. Phys.}\ }\textbf {\bibinfo
  {volume} {57}},\ \bibinfo {pages} {353} (\bibinfo {year} {2008})}\BibitemShut
  {NoStop}%
\bibitem [{\citenamefont {Stadler}\ \emph {et~al.}(2012)\citenamefont
  {Stadler}, \citenamefont {Krinner}, \citenamefont {Meineke}, \citenamefont
  {Brantut},\ and\ \citenamefont {Esslinger}}]{esslinger2012}%
  \BibitemOpen
  \bibfield  {author} {\bibinfo {author} {\bibfnamefont {D.}~\bibnamefont
  {Stadler}}, \bibinfo {author} {\bibfnamefont {S.}~\bibnamefont {Krinner}},
  \bibinfo {author} {\bibfnamefont {J.}~\bibnamefont {Meineke}}, \bibinfo
  {author} {\bibfnamefont {J.-P.}\ \bibnamefont {Brantut}}, \ and\ \bibinfo
  {author} {\bibfnamefont {T.}~\bibnamefont {Esslinger}},\ }\href@noop {}
  {\bibfield  {journal} {\bibinfo  {journal} {Nature}\ }\textbf {\bibinfo
  {volume} {491}},\ \bibinfo {pages} {736} (\bibinfo {year}
  {2012})}\BibitemShut {NoStop}%
\bibitem [{\citenamefont {Duarte}\ \emph {et~al.}(2011)\citenamefont {Duarte},
  \citenamefont {Hart}, \citenamefont {Hitchcock}, \citenamefont {Corcovilos},
  \citenamefont {Yang}, \citenamefont {Reed},\ and\ \citenamefont
  {Hulet}}]{hulet2011}%
  \BibitemOpen
  \bibfield  {author} {\bibinfo {author} {\bibfnamefont {P.~M.}\ \bibnamefont
  {Duarte}}, \bibinfo {author} {\bibfnamefont {R.~A.}\ \bibnamefont {Hart}},
  \bibinfo {author} {\bibfnamefont {J.~M.}\ \bibnamefont {Hitchcock}}, \bibinfo
  {author} {\bibfnamefont {T.~A.}\ \bibnamefont {Corcovilos}}, \bibinfo
  {author} {\bibfnamefont {T.-L.}\ \bibnamefont {Yang}}, \bibinfo {author}
  {\bibfnamefont {A.}~\bibnamefont {Reed}}, \ and\ \bibinfo {author}
  {\bibfnamefont {R.~G.}\ \bibnamefont {Hulet}},\ }\href@noop {} {\bibfield
  {journal} {\bibinfo  {journal} {Phys. Rev. A}\ }\textbf {\bibinfo {volume}
  {84}},\ \bibinfo {pages} {061406} (\bibinfo {year} {2011})}\BibitemShut
  {NoStop}%
\bibitem [{\citenamefont {Hsiao}\ \emph {et~al.}(2014)\citenamefont {Hsiao},
  \citenamefont {Chen}, \citenamefont {Tsai},\ and\ \citenamefont
  {Chen}}]{hsiao2014}%
  \BibitemOpen
  \bibfield  {author} {\bibinfo {author} {\bibfnamefont {Y.-F.}\ \bibnamefont
  {Hsiao}}, \bibinfo {author} {\bibfnamefont {H.-S.}\ \bibnamefont {Chen}},
  \bibinfo {author} {\bibfnamefont {P.-J.}\ \bibnamefont {Tsai}}, \ and\
  \bibinfo {author} {\bibfnamefont {Y.-C.}\ \bibnamefont {Chen}},\ }\href@noop
  {} {\bibfield  {journal} {\bibinfo  {journal} {Phys. Rev. A}\ }\textbf
  {\bibinfo {volume} {90}},\ \bibinfo {pages} {055401} (\bibinfo {year}
  {2014})}\BibitemShut {NoStop}%
\bibitem [{\citenamefont {Blatt}\ \emph {et~al.}(2014)\citenamefont {Blatt},
  \citenamefont {Halfmann},\ and\ \citenamefont {Peters}}]{blatt2014}%
  \BibitemOpen
  \bibfield  {author} {\bibinfo {author} {\bibfnamefont {F.}~\bibnamefont
  {Blatt}}, \bibinfo {author} {\bibfnamefont {T.}~\bibnamefont {Halfmann}}, \
  and\ \bibinfo {author} {\bibfnamefont {T.}~\bibnamefont {Peters}},\
  }\href@noop {} {\bibfield  {journal} {\bibinfo  {journal} {Opt. Lett.}\
  }\textbf {\bibinfo {volume} {39}},\ \bibinfo {pages} {446} (\bibinfo {year}
  {2014})}\BibitemShut {NoStop}%
\bibitem [{\citenamefont {Keitel}\ \emph {et~al.}(1992)\citenamefont {Keitel},
  \citenamefont {Scully},\ and\ \citenamefont {S\"ussmann}}]{keitel1992}%
  \BibitemOpen
  \bibfield  {author} {\bibinfo {author} {\bibfnamefont {C.~H.}\ \bibnamefont
  {Keitel}}, \bibinfo {author} {\bibfnamefont {M.~O.}\ \bibnamefont {Scully}},
  \ and\ \bibinfo {author} {\bibfnamefont {G.}~\bibnamefont {S\"ussmann}},\
  }\href@noop {} {\bibfield  {journal} {\bibinfo  {journal} {Phys. Rev. A}\
  }\textbf {\bibinfo {volume} {45}},\ \bibinfo {pages} {3242} (\bibinfo {year}
  {1992})}\BibitemShut {NoStop}%
\bibitem [{\citenamefont {Rohringer}\ \emph {et~al.}(2012)\citenamefont
  {Rohringer}, \citenamefont {Ryan}, \citenamefont {London}, \citenamefont
  {Purvis}, \citenamefont {Albert}, \citenamefont {Dunn}, \citenamefont
  {Bozek}, \citenamefont {Bostedt}, \citenamefont {Graf}, \citenamefont {Hill}
  \emph {et~al.}}]{rohringer2012}%
  \BibitemOpen
  \bibfield  {author} {\bibinfo {author} {\bibfnamefont {N.}~\bibnamefont
  {Rohringer}}, \bibinfo {author} {\bibfnamefont {D.}~\bibnamefont {Ryan}},
  \bibinfo {author} {\bibfnamefont {R.}~\bibnamefont {London}}, \bibinfo
  {author} {\bibfnamefont {M.}~\bibnamefont {Purvis}}, \bibinfo {author}
  {\bibfnamefont {F.}~\bibnamefont {Albert}}, \bibinfo {author} {\bibfnamefont
  {J.}~\bibnamefont {Dunn}}, \bibinfo {author} {\bibfnamefont {J.}~\bibnamefont
  {Bozek}}, \bibinfo {author} {\bibfnamefont {C.}~\bibnamefont {Bostedt}},
  \bibinfo {author} {\bibfnamefont {A.}~\bibnamefont {Graf}}, \bibinfo {author}
  {\bibfnamefont {R.}~\bibnamefont {Hill}},  \emph {et~al.},\ }\href@noop {}
  {\bibfield  {journal} {\bibinfo  {journal} {Nature}\ }\textbf {\bibinfo
  {volume} {481}},\ \bibinfo {pages} {488} (\bibinfo {year}
  {2012})}\BibitemShut {NoStop}%
\bibitem [{\citenamefont {Cavaletto}\ \emph {et~al.}(2014)\citenamefont
  {Cavaletto}, \citenamefont {Harman}, \citenamefont {Ott}, \citenamefont
  {Buth}, \citenamefont {Pfeifer},\ and\ \citenamefont
  {Keitel}}]{cavaletto2014}%
  \BibitemOpen
  \bibfield  {author} {\bibinfo {author} {\bibfnamefont {S.~M.}\ \bibnamefont
  {Cavaletto}}, \bibinfo {author} {\bibfnamefont {Z.}~\bibnamefont {Harman}},
  \bibinfo {author} {\bibfnamefont {C.}~\bibnamefont {Ott}}, \bibinfo {author}
  {\bibfnamefont {C.}~\bibnamefont {Buth}}, \bibinfo {author} {\bibfnamefont
  {T.}~\bibnamefont {Pfeifer}}, \ and\ \bibinfo {author} {\bibfnamefont
  {C.~H.}\ \bibnamefont {Keitel}},\ }\href@noop {} {\bibfield  {journal}
  {\bibinfo  {journal} {Nature Photonics}\ }\textbf {\bibinfo {volume} {8}},\
  \bibinfo {pages} {520} (\bibinfo {year} {2014})}\BibitemShut {NoStop}%
\bibitem [{\citenamefont {Wang}\ \emph {et~al.}(2015)\citenamefont {Wang},
  \citenamefont {Zhu}, \citenamefont {Evers},\ and\ \citenamefont
  {Scully}}]{wang2015}%
  \BibitemOpen
  \bibfield  {author} {\bibinfo {author} {\bibfnamefont {D.-W.}\ \bibnamefont
  {Wang}}, \bibinfo {author} {\bibfnamefont {S.-Y.}\ \bibnamefont {Zhu}},
  \bibinfo {author} {\bibfnamefont {J.}~\bibnamefont {Evers}}, \ and\ \bibinfo
  {author} {\bibfnamefont {M.~O.}\ \bibnamefont {Scully}},\ }\href@noop {}
  {\bibfield  {journal} {\bibinfo  {journal} {Phys. Rev. A}\ }\textbf {\bibinfo
  {volume} {91}},\ \bibinfo {pages} {011801} (\bibinfo {year}
  {2015})}\BibitemShut {NoStop}%
\end{thebibliography}
%


\pagebreak
\widetext
\begin{center}
\textbf{\large Supplemental Materials: All-optical control of superradiance and matter waves using a dispersive cavity}

{\normalsize Shih-Wei Su, Zhen-Kai Lu, Nina Rohringer, Shih-Chuan Gou and Wen-Te Liao}

(Dated: \today)
\end{center}
\setcounter{equation}{0}
\setcounter{figure}{0}
\setcounter{table}{0}
\setcounter{page}{1}
\makeatletter
\renewcommand{\theequation}{S\arabic{equation}}
\renewcommand{\thefigure}{S\arabic{figure}}
\renewcommand{\bibnumfmt}[1]{[S#1]}
\renewcommand{\citenumfont}[1]{S#1}
In this supplementary material, we present the equation of motion and the condensate dynamics in more detail.  

In the presence of counter-propagating coupling fields $\Omega_c^{\pm}$ and the superradiant fields $\Omega^{\pm}$, the dynamics of the EIT mirror is described by the optical-Bloch equation which reads:
\begin{eqnarray}
\partial_{t}\rho_{12} & = & -i\left(\Delta-\Delta_{c}\right)\rho_{12}-i(\Omega_{c}^{+}\rho_{13}^{+}+\Omega_{c}^{-}\rho_{13}^{-})/2+i(\Omega^{+*}\rho_{32}^{+}+\Omega^{-*}\rho_{32}^{-})/2,\label{eq:rho12-SLP}
\end{eqnarray}
\begin{eqnarray}
\partial_{t}\rho_{13}^{\pm} & = & -\left(i\Delta+\Gamma/2\right)\rho_{13}^{\pm}-i\Omega_{c}^{\pm*}\rho_{12}/2-i\Omega^{\pm*}\left(\rho_{11}-\rho_{33}\right)/2,\label{eq:rho13p-SLP}
\end{eqnarray}
\begin{eqnarray}
\partial_{t}\rho_{23}^{\pm} & = & -\left(i\Delta_{c}+\Gamma/2\right)\rho_{23}^{\pm}-i\Omega^{\pm*}\rho_{21}+i\Omega_{c}^{\pm*}\left(\rho_{33}-\rho_{22}\right),\label{eq:rho23-SLP}
\end{eqnarray}
\begin{eqnarray}
\partial_{t}\rho_{11}=\Gamma\rho_{33}/2+Im\left[\Omega^{+}\rho_{13}^{+}+\Omega^{-}\rho_{13}^{-}\right],\label{eq:rho11-SLP}
\end{eqnarray}
\begin{eqnarray}
\partial_{t}\rho_{22}=\Gamma\rho_{33}/2+Im\left[\Omega_{c}^{+}\rho_{23}^{+}+\Omega_{c}^{-}\rho_{23}^{-}\right],\label{eq:rho22-SLP}
\end{eqnarray}
\begin{eqnarray}
\partial_{t}\rho_{33} & = & -\Gamma\rho_{33}-Im\left[\Omega^{+}\rho_{13}^{+}+\Omega^{-}\rho_{13}^{-}+\Omega_{c}^{+}\rho_{23}^{+}+\Omega_{c}^{-}\rho_{23}^{-}\right],\label{eq:rho33-SLP}
\end{eqnarray}
where $\Gamma$ is the spontaneous decay rate of the excited state
$\left|3\right\rangle $ and $\rho_{ij}$ is the element of the density
matrix. In equations~(\ref{eq:rho12-SLP})-(\ref{eq:rho33-SLP}), all fast-oscillation exponential factors associating with center frequencies and wave factors have been eliminated, and only slowly-varying profiles are retained. 

\begin{figure*}[bht]
\includegraphics[width=0.7\paperwidth]{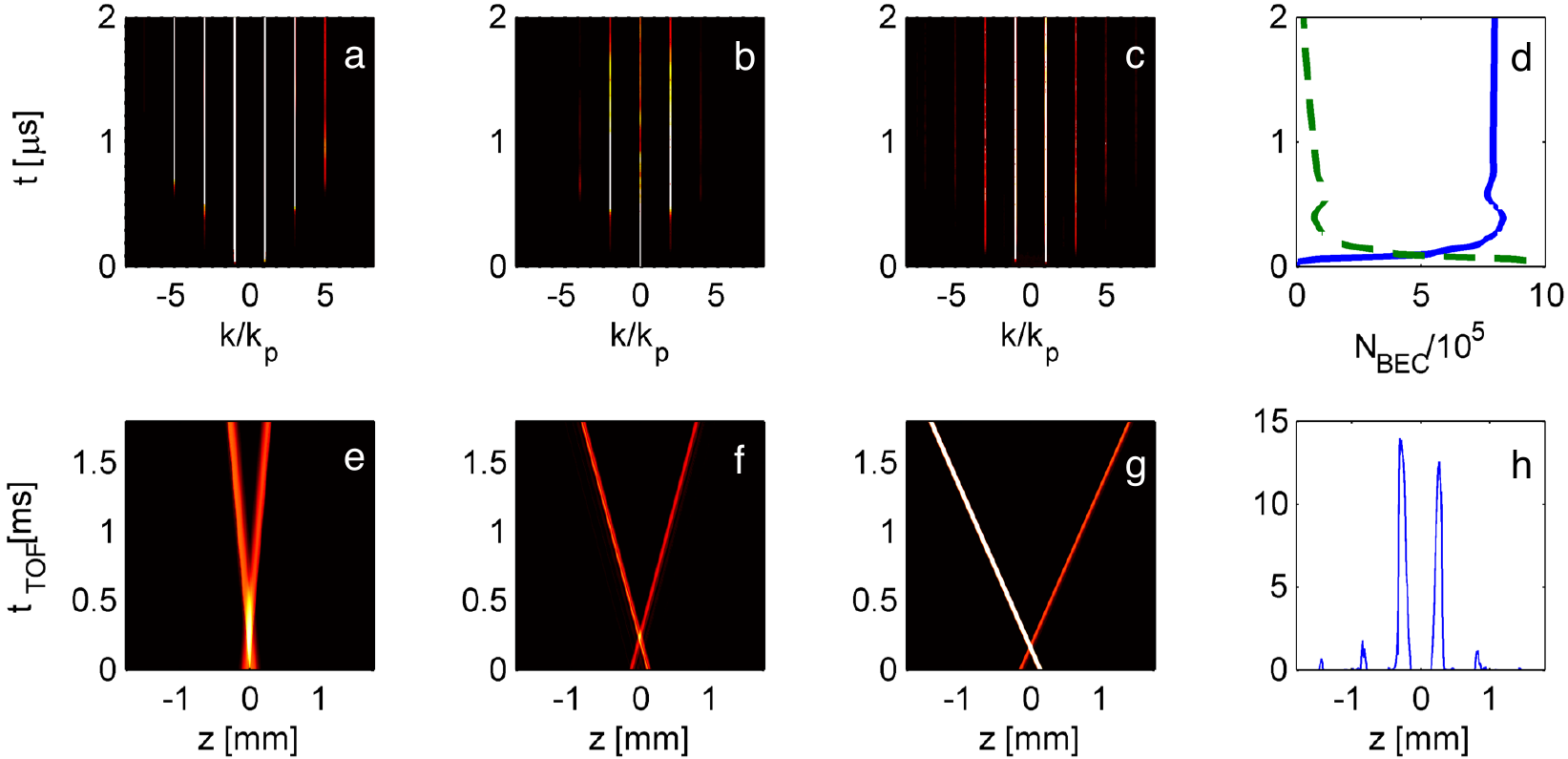}
\caption{Panels \textbf{a}-\textbf{d} show the dynamics of the binary BEC inside the EIT
cavity of $d^{opt}=500$ and $\Omega_{c}=2\Gamma$ within $2$ $\mu$s. The evolution of the condensate densities
in momentum space are shown in panels \textbf{a} and \textbf{b} where the high odd
$k_{p}$ modes are generated in $\psi_{g}$ while the even $k_{p}$
modes are generated in $\psi_{e}$. Panel \textbf{c} shows the optical coherence
$\sigma_{eg}$. The coherent transfer of the BEC particle number from states
$\left|e\right\rangle $ (green dashed-dotted line) to $\left|g\right\rangle $
(blue solid line) is shown. Panels \textbf{e}-\textbf{h} represent the time-of-fight
simulation after the condensate is released from the cavity and harmonic
trap. Panels \textbf{e}-\textbf{g} are the density components carry the momenta
$\pm k_{p}$, $\pm3k_{p}$, and $\pm5k_{p}$, respectively. The total
density profile $\left|\psi_{g}\right|^{2}$at $t_{TOF}=1.8$ ms.
The contrast of the 2D plots is adjusted for better visualization.}
\label{single-trajectory-r0-3um-OD-500-Om-2} 
\end{figure*}

When the condensate is loaded in the dispersive cavity, in the presence of the interaction between the circulating SR fields and condensate, the ground-state and excited condensate wave functions constitute the superposition of discrete $\pm (2n+1)k_p$ and $\pm 2nk_p$ plane waves, respectively. In this manner the condensate wave function can be decomposed as : 
\begin{align}
\psi_g=\sum_{n=-\infty}^{\infty}\psi_{g}^{(n)}e^{i(2n+1)k_{p}z},\label{eq:ground-state wave}
\end{align}
and
\begin{align*}
\psi_e=\sum_{n=-\infty}^{\infty}\psi_{e}^{(n)}e^{i2nk_{p}z},
\end{align*}
where $\psi_{e,g}^{(n)}$ are slowly-varying profiles.  As depicted in Fig.~\ref{single-trajectory-r0-3um-OD-500-Om-2} \textbf{a} and \textbf{b}, the momentum-space density profile of single realization of $d^{opt}=500$ and $\Omega_{c}=2\Gamma$ shows the clear generation of these discrete $\pm nk_{p}$ modes. Furthermore the Fourier transform of the coherence $\sigma_{eg}$ depicted in Fig.~\ref{single-trajectory-r0-3um-OD-500-Om-2} \textbf{c} shows clear superposition of $\pm nk_p$ modes.

The coherent SR-BEC interaction transfer most of the atoms to the ground state while the rest of them decay to other states due to the incoherent spontaneous processes as shown in Fig.~\ref{single-trajectory-r0-3um-OD-500-Om-2} \textbf{d}.  Therefore
the dynamics of the condensate after passing through the cavity is described by the single component Gross-Pitaevskii equation which reads

\begin{align}
i\hbar\partial_{t}\psi_{g} & =\left(-\frac{\hbar^{2}}{2m}\partial_{z}^{2}+\frac{m\omega_{z}^2z^2}{2}+g_{gg}\left|\psi_{g}\right|^{2}\right)\psi_{g},\label{eq:GPE-ground}
\end{align}
where we neglect the cross-species and atom-light  interactions.

During the time of flight (TOF) measurement, the condensate would split into several atomic clouds which corresponds to different $\pm nk_p$ modes. To simulate the TOF dynamics, we numerically integrated equation.~\ref{eq:GPE-ground} by removing the trapping potential and interactions.
As shown in Fig.~\ref{single-trajectory-r0-3um-OD-500-Om-2} \textbf{e}-\textbf{h},
we perform the TOF simulation for $\Delta t_{TOF}=1.8$ ms.
In Fig.~\ref{single-trajectory-r0-3um-OD-500-Om-2} \textbf{e}-\textbf{g}, the evolution
of the condensed atoms carrying momenta are plotted and the velocities
can be calculated from the slopes which agrees with the expected values,
$\pm\hbar k_{p}/m$, $\pm3\hbar k_{p}/m$, and $\pm\hbar k_{p}/m$.
In Fig.~\ref{single-trajectory-r0-3um-OD-500-Om-2} \textbf{h}, the total
density distribution at $t_{TOF}=1.8$ ms is shown where six density bumps that are
symmetric to the origin can be clearly observed. The two innermost density
bump pair in Fig.~\ref{single-trajectory-r0-3um-OD-500-Om-2} \textbf{h} corresponds
to the $\pm k_{p}$ modes while the density bump pair located around $z=\pm 0.5$ mm is the $\pm3k_{p}$
modes and the outermost pair carries $\pm5k_{p}$ momenta.

\end{document}